\documentstyle{mn}

\title[The Proximity Effect in QSO pairs]
{The Proximity Effect on the Lyman $\alpha$ Forest due to a foreground QSO}

\author[A. Fern\'andez-Soto et al.]
       {A. Fern\'andez-Soto$^1$, X. Barcons$^2$, R. Carballo$^1$, J.K. Webb$^3$
	\\
	$^1$ Departamento de F\'{\i}sica Moderna, Universidad de Cantabria.
	39005, Santander, Spain \\
	$^2$ Instituto de F\'{\i}sica de Cantabria.\\
	Consejo Superior de Investigaciones
	Cient\'{\i}ficas - Universidad de Cantabria. 39005, Santander, Spain \\
	$^3$ Kensington School of Physics. University of New South Wales.
	Sidney, Australia.}

\date{Accepted 1995 June 7. Received 1995 February 27; in original form 1994
October 11}

\newcommand{\dndz}{\frac{d{\cal N}}{dz}}

\newcommand{\erg}{{\rm erg}\, {\rm cm}^{-2}\, {\rm s}^{-1}\, {\rm Hz}^{-1}\,
{\rm srad}^{-1}}
\newcommand{\lya}{{Lyman $\alpha$}}
\newcommand{\lyb}{{Lyman $\beta$}}
\newcommand{\kms}{{\rm km\, s}^{-1}}
\newcommand{\pcm}{{\rm cm}^{-2}}

\newcommand{\gauss}{\frac{1}{\sigma_{b} \sqrt{2\pi}} \exp
\left(- \frac{(b-\overline{b})^{2}}{2\sigma_{b}^{2}} \right)}

\begin{document}

\maketitle

\begin{abstract}

The influence of a foreground QSO on the \lya\ forest of another
QSO with higher redshift has been investigated by analyzing the spectra of
three such objects at redshifts $z=2 - 2.7$.  This influence is not
contaminated by any projection effects, as opposed to the inverse
effect along the line of sight, where incomplete coverage of the QSO
continuum emitting region by the \lya\ clouds could contribute to the
relative lack of lines.  Our results are consistent with the existence
of a proximity effect due to the foreground QSO, but due to its
weakness we can only reject the absence of such effect at $\sim
1\sigma$ level. By modelling this proximity effect assuming that \lya\
clouds are low-density highly ionized objects we find that the best
value for the UV photoinizing intensity at those redshifts is $\sim
10^{-20.5}\, \erg$ at the Lyman limit which is consistent with
previous estimates of the background based on the inverse effect. We
also find an absolute lower limit (at 95 per cent confidence) to the UV
intensity at the level of $10^{-21.8}\, \erg$ which means the rejection of
a number of models for the UV background where it is mostly contributed by
QSOs, and absorption by Lyman limit systems is taken into account.

\end{abstract}


\section{INTRODUCTION}

The inverse effect in the redshift distribution of the \lya\ absorption lines
seen towards distant QSOs (which is a relative lack of these lines at redshifts
close to the QSO one) was first detected by Weyman, Carswell \& Smith (1981).
It was soon realized that different values were obtained for the evolution
parameter $\gamma$ when trying to fit a power law to the number density of
lines per unit redshift,
\begin{equation}
{d{\cal N}\over dz}=A\, (1+z)^{\gamma}
\end{equation}
depending on whether a single QSO was taken or a sample of them was used. The
fit of such form to any single QSO always resulted in lower values of  $\gamma$
than when several QSO samples were considered as a whole  (Carswell et al.
1982). Murdoch et al. (1986) explored this effect with a large  sample of \lya\
lines obtained from intermediate- to low- resolution spectroscopy  of several
QSOs, concluding that there was a relative decrease in the line  number density
when the absorption redshift approached the emission redshift.  Tytler (1987),
by using a larger but more inhomogeneous sample, reached a  similar conclusion,
except for the fact that he claimed that the inverse effect  (or anomaly) was
detected far away from the observed QSO, whilst Murdoch et al  (1986) concluded
that this effect was restricted to a distance $\sim 8\,  h_{50}^{-1}$ from the
QSO.  Webb \& Larsen (1988), in a detailed analysis of a large sample,   also
concluded that the inverse effect was concentrated at a short distance from the
observed QSO.

Several models have been suggested to explain the inverse effect. The first
one invokes some sort of local mechanism  around the observed QSO that renders
undetectable the \lya\ clouds close to it, for example in terms of enhanced
photoionization by the QSO (Murdoch et al. 1986; Tytler 1987; Bajtlik, Duncan
\& Ostriker 1988, hereafter BDO).  Another class of models postulates a
rather small size for the \lya\ clouds, in such a way that at redshifts close
to the QSO they are unable to cover its continuum emitting region, resulting
in a relative lack of lines above any given threshold (Tytler 1987). Finally
Lu, Wolfe \& Turnshek (1991) tested whether a flattening of the power law (1)
at high redshift could explain the inverse effect.  Although these authors
find some evidence for this flattening, they conclude that it does not remove
the inverse effect at all.

If the photoionization explanation for the inverse effect is to be assumed
real, the \lya\ clouds would be galaxy-sized objects, with a rather low density
$n_{H}< 10^{-4}\, {\rm cm}^{-3}$ and a high ionization fraction
$n_{HII} / n_{HI} \sim 10^{5}$.  Collisional ionization would then be
negligible and the optically thin gas would therefore be mostly ionized by the
UV photons from the general background plus the ones contributed by any nearby
QSO.  If, on the contrary, the reason for the inverse effect is the small size
of the \lya\ clouds, these objects would have large densities ($n_{H}>1\, {\rm
cm}^{-3}$), they would be almost neutral (except for a small fraction of
collisionally ionized atoms) and photoionization by UV photons would be less
important.  All the models described above present some problems. For the
second class of models the \lya\ clouds should be very small ($\sim pc$ size)
in contradiction with the compelling evidence that they have a size similar to
a galaxy  (Hippelein \& Meisenheimer 1993) or even larger (Smette et al. 1992;
Dinshaw et al. 1994; Bechtold et al. 1994). These models also require an
unphysical fine tuning of the cloud size and the QSO continuum emitting region
(Barcons \& Fabian 1987). On the other hand, the photoionization model predicts
a correlation between the magnitude of the inverse effect and  the ionizing
flux from the QSO, a fact whose observational confirmation is still unclear
(Barcons \& Fabian 1987; Webb \&Larsen 1988; Lu et al. 1991, Bechtold 1994).

Therefore, the present situation is that the most likely explanation for the
inverse effect is that the enhanced photoionization field around the QSO
evaporates or overionizes the surrounding clouds.

In order to provide a complementary test for the local (photoionization)
models, we have searched for the possible effect introduced in the \lya\ forest
of a background QSO by a foreground QSO which lies close to the line of sight
under study.  If the inverse effect is due to the photoionization from the QSO,
in our situation, the foreground object should also influence the \lya\ forest
of the background QSO.  If, on the contrary, the small size of the clouds is
the reason for the inverse effect, we should not detect any influence from the
foreground QSO. Similar works have been presented previously by Crotts (1989)
and Dobrzycki \& Bechtold (1991).

After having studied three such cases, we have only been able to marginally
detect what we shall call the {\it proximity effect} (following the suggestion
by BDO) at  $\sim 1\sigma$ confidence. Subsequent modelling of this effect
enables us to estimate the general UV ionizing background for which we have a
result entirely consistent with estimates based on the inverse effect itself.
Under the photoionization scheme, we find firm lower limits to the UV ionizing
background  which enable us to reject some of the models that have been put
forward to account for the high-z UV intensity.

The paper is organised as follows.  Section 2 presents the objects
selected for our study and a summary of the observations.  In Sections
3 and 4 we discuss with detail the absorption line lists obtained by
line fitting the spectra of the three objects used. Much attention is
devoted to identifying the heavy-element systems (Section 3) in order to
remove any lines in the \lya\ forest that do not correspond to \lya\
clouds.  Some discussion on the distribution of the clouds is also
presented based on our data in Section 4. In Section 5, we explain and
perform the test for the proximity effect after modelling it in terms
of a simple photoionization model.  Section 6 presents a brief
discussion of our findings, and implications for the origin of the high
redshift UV background.

\section{THE DATA}

\subsection{The Sample}

A search has been made in the Hewitt and Burbidge Catalogue
(Hewitt \& Burbidge 1993) for every pair of QSOs matching the
following conditions:

\begin{enumerate}[(4)]
 \renewcommand{\theenumi}{(\roman{enumi})}

\item The background QSO (redshift $z_b$) must have a significant part (at
least $40000\, \kms$) of its \lya\  forest (i.e., the spectral region contained
between the \lya\ and \lyb\ emission lines) in the visible region of the
spectrum ($\lambda>3250$~\AA ).

\item The foreground QSO must have a redshift ($z_f$) such that it is able to
influence the visible  fraction of the \lya\ forest of the background QSO,
i.e., $1216\, (1+z_f)$~\AA\ must fall within the observable \lya\ forest of the
background object {\it and} be further than $5000\, \kms$ from $1216\,
(1+z_b)$~\AA\ in order to avoid overlapping of the inverse effect and the
proximity effect we are looking for.

\item The magnitude of the background QSO should be $\leq$~18, so that
spectroscopy at moderate resolution can be easily obtained.

\item The angular separation between both QSOs should be less than 20 arcmin,
which corresponds to \linebreak $\sim 10\, h_{50}\, {\rm Mpc}$ at
$z\sim 2.5$. Here, and throughout this paper, we use the standard cosmology
parameters $q_0=0.5$ and $H_0=50\, h_{50}\, \kms\, {\rm Mpc}^{-1}$.

\end{enumerate}

The selection process produced a preliminary sample containing about 30 such
pairs. After another filtering process, based on use of other catalogues to
avoid misclassified objects and possibilities of observation from the
Observatorio del Roque de los Muchachos at La Palma (Canary Islands, Spain),
we obtained a reduced target sample with  10 QSO pairs.

% Including table 1 here

\begin{table}
  \begin{minipage}{83mm}
  \caption{The observed QSO pairs.}
  \label{tab-sample}
  \begin{tabular}{ l c c c c c }\hline
 OBJECT & \multicolumn{2}{c}{COORDINATES}\footnote{J2000}& $m_{V}$ & z
& $\theta$\footnote{Angular separation between lines of sight, in arcmin}\\
& $\alpha$ & $\delta$ &&&\\
 1055+021 & 10 57 57 &  01 54 03  & 17.8&2.73 &          \\
 1055+021 & 10 57 13 &  01 47 55  & 20.0&2.29 & 12.6     \\
&&&&&\\
 1222+228 & 12 25 27 &  22 35 13  & 15.5&2.051&          \\
 1222+228 & 12 25 24 &  22 31 28  & 19.0&1.87 &  3.8     \\
&&&&&\\
 1228+077 & 12 31 21 &  07 25 18  & 17.6&2.391&          \\
 1228+077 & 12 31 08 &  07 24 39  & 17.5&1.878&  5.7     \\ \hline
  \end{tabular}
\end{minipage}
\end{table}

All through this paper, each pair will be named according to the name of the
background QSO.  Two of our pairs (1307+296 and 1305+298) turned up to be
stars, as was seen by taking short-exposure spectra at low resolution with
the William Herschel Telescope in 1993 February.

The sample that we observed, consisting of three QSOs is shown in Table 1.
These QSOs span a rather restricted redshift range ($z=2 - 2.7$) but cover a
wider range in terms of angular separations from the foreground QSO
($\theta = 3.8 - 12.6$). That will allow us to test the existence of the
proximity effect under different circumstances (see Section 5).

\subsection{Observations and data reduction}

All observations were performed at the Observatorio del Roque de los Muchachos
in the island of La Palma (Canary Islands, Spain). Data from QSO1228+077 were
acquired using the Isaac Newton Telescope with the  IDS spectrograph and IPCS
(Image Photon Counting System) detector in 1989 February. The data from
QSO1222+228  were taken in 1993 February, using the William Herschel Telescope
with the ISIS double spectrograph and the IPCS-II as detector in the blue arm.
Finally, data from  QSO1055+021 were taken in 1994 February, again using the
ISIS double spectrograph and the TEK/CCD detector on the blue arm of the same
telescope. A reduced version of the observing logs is listed in Table 2.

% Including table 2 here

\begin{table*}
 \centering
  \begin{minipage}{170mm}

\begin{small}

  \caption{Summary of observations.}
  \label{observing-log}
  \begin{tabular}{ l c c c c c c c c c } \hline
 OBJECT   & INSTRUMENT & DATE & WAVELENGTH
& EXP.       & SEEING     & $\sigma_v$\footnote{Dispersion of the spectrum.}
& S/N\footnote{Minimum and maximum signal-to-noise ratio as estimated from the
spectra.} \\
          &            &      & RANGE (\AA )
&  TIME (h)  &   (arcsec) & ($\kms$) & \\
 1055+021 & WHT+ISIS+TEK/CCD & 13-Feb-94 & 3910 - 4111 & 3.3
& 1     & 13.2 & 9-13 \\
&&&&&&&\\
 1222+228 & WHT + ISIS + IPCS-II & 14-Feb-93 & 3255 - 3695 &  1.1
& 1.5   &      &        \\
 1222+228 & WHT + ISIS + IPCS-II & 15-Feb-93 & 3255 - 3695 &  1.7
& 2     & 36.3 & 6-12 \\
&&&&&&&\\
 1228+077 & INT + IDS  + IPCS    &  8-Feb-89 & 3360 - 3845 &  2.8
& 1.5   &      &        \\
 1228+077 & INT + IDS  + IPCS    &  9-Feb-89 & 3360 - 3845 &  5.0
& 1.5   & 34.1 & 4-7  \\ \hline
  \end{tabular}

\end{small}

  \end{minipage}
\end{table*}

The choice of the IPCS and IPCS-II to observe QSO1228+077 and QSO1222+228 was
done because the crucial part of the spectrum (i.e., the part putatively
influenced by the foreground QSO) falls at $\lambda<4000$~\AA\, where this
detector has better sensitivity than any of the available CCDs at La Palma.

The blue gratings were chosen in order to have an approximate reciprocal
resolution of about $\sim 0.2\, {\rm \AA}/{\rm pixel}$. This produced a
spectral resolution good enough to allow profile fitting of the absorption
lines present in the different spectra.

Although we tuned the red arm settings of the ISIS double spectrograph  (during
the observations of QSO1222+228 and QSO1055+021), in order to identify metal
lines of possible heavy-element absorption systems with associated \lya\ lines
falling in the blue arm region, we did not find any line that either confirmed
or rejected any possible system.

The data reduction  was carried out using the Starlink FIGARO package. We used
standard techniques, except that we needed to generate and propagate
through the entire process an error array for which we had to make the suitable
modifications to the standard routines provided by FIGARO. To obtain a precise
wavelength-scale, calibration arc lamps were exposed before and after each
integration on all of our objects. Heliocentric and air-to-vacuum corrections
were performed. No attempt has been made to obtain a flux callibration for
any of the QSOs, which does not affect to the parameters of the absorption
lines.

As it can be seen from Table 2, the data from QSO1055+021 are of a much higher
quality, caused by the longer exposure time and better sky conditions during
their acquisition. None the less, data from all three QSOs have spectral
resolution and signal-to-noise ratio good enough to perform absorption line
fitting.

Continuum fitting was carried out using a standard spline method. Selected
regions of each spectrum were used in an iterative way that rejected in each
iteration the points that were significantly deviant  from the fitted
continuum, until the fitting procedure converged.

A preliminary absorption line list was obtained searching for significant
(greater than 4$\sigma$) deviations from the continuum in the normalized
spectra, following the method presented in Young et al. (1979).

Voigt profiles were fitted to each line (or blend of lines)  in the spectra
using the absorption line fitting code ALF developed by ourselves. This program
performs $\chi^2$ minimisation of a spectral region  by introducing a Voigt
profile for each line (that is why the error array is needed). The program
finds the best-fitting column density, Doppler dispersion parameter and
redshift for each one of the lines, assuming that they correspond to given
ions and atomic transitions. ALF also produces an estimate of the errors on
these parameters and a goodness-of-fit given the minimum $\chi^2$ reached and
the number of degrees of freedom.   A fit was considered to be good enough if
it is not rejected by more than 99 per cent probability, according to $\chi^2$
statistics. If the fit over a selected region was not good enough, another
line (with free column density, velocity dispersion parameter and redshift)
was added and the fit repeated.  The process was stopped either when the
rejection probability is less than 99 per cent and the fit looked good to the
eye, or when addition of a new line resulted in an {\it increase} of the
rejection probability.  This avoids overfitting to a large extent. The
resulting line lists are presented in Table 3 (QSO1055+021), Table 4
(QSO1222+228) and Table 5 (QSO1228+077).

% Insertar Figuras 1,2 y 3, cada una en una pagina y en landscape
%
%\begin{figure*}
%  \vbox to 200mm{\vfil}
%  \caption{}
%  \label{spec-1055}
%\end{figure*}
%
%\begin{figure*}
%  \vbox to 200mm{\vfil}
%  \caption{}
%  \label{spec-1222}
%\end{figure*}
%
%\begin{figure*}
%  \vbox to 200mm{\vfil}
%  \caption{}
%  \label{spec-1228}
%\end{figure*}

In Figures 1,2 and 3 we show the spectra of the 3 objects QSO1055+021,
QSO1222+228 and QSO1228+077 respectively, with arrows marking the wavelength
of the Lyman $\alpha$ emission line for each respective foreground QSO.
These spectra have been normalized to the fitted continuum. The spectral fit
achieved is also shown as a dotted line. There are several features in these
spectra that we do not believe to be absorption lines, since their width is
too small for the spectral resolution we have obtained.  In fact we tried to
fit these `lines' and found that the absence of a line provided a better fit
than the presence of an absorption line. The most conspicuous of these
features, which we believe to be spikes of the noise, appear around 4010 and
4108 \AA\ in Fig. 1, and in 3487 and in 3762 \AA\ in Fig. 3.

% Insertar Tablas 3, 4 y 5
%\begin{table*}
%  \vbox to 200mm{\vfil}
%  \caption{}
%  \label{tab-1055}
%\end{table*}
%
%\begin{table*}
%  \vbox to 200mm{\vfil}
%  \caption{}
%  \label{tab-1055bis}
%\end{table*}
%
%\begin{table*}
%  \vbox to 200mm{\vfil}
%  \caption{}
%  \label{tab-1222}
%\end{table*}
%
%\begin{table*}
%  \vbox to 200mm{\vfil}
%  \caption*{}
%  \label{tab-1222bis}
%\end{table*}

\section{The Heavy-element absorption systems}

A fraction of the absorption lines observed in QSO spectra are supposed to
be caused by absorption from elements other than HI. These are the so-called
metal lines, which have their origin in systems rich in metals and are
called heavy-element absorption systems.

In order to search for heavy-element absorption systems in our list, we
cross-correlated it with a list of the most relevant atomic transitions
following the method by Young et al. (1979). We used the same `short list'
as in their paper, with some additions in order to make it denser in certain
wavelength ranges. The new added lines are CI $\lambda 1261$, AlII $\lambda
1670$, AlIII  $\lambda\lambda 1854,1862$, MgI $\lambda 2026$, BeII $\lambda
3131$, NaI $\lambda 3303$, AlI $\lambda\lambda 3083,3093$ and CaI $\lambda
4227$.

As a result of this correlation analysis, a function representing the number
of coincidences of different lines with the same absorption system redshift
is obtained.  The peaks of this cross-correlation function are prime
candidates for correspondance to heavy-element systems. In order to assess
the significance of  these peaks, we generated 100 synthetic line lists
uniformly distributed over the available wavelength range for each spectrum,
and performed the same cross-correlation analysis. Averaging of these 100
cross-correlation functions provided a measure of the noise level for the
function corresponding to the real data. All the peaks of the
cross-correlation function above $1\sigma$ were carefully studied using the
complete line list obtained by the line fitting procedure rather than from
the preliminary list produced by FIGRED. Special attention was payed to
multiplet transitions of MgII, CIV, SiII, SiIV and AlIII. We considered that
a putative heavy-element system is real if it  matches the following
conditions:

\begin{enumerate}[(3)]
 \renewcommand{\theenumi}{(\roman{enumi})}

\item A system with $\lambda_{min} \leq 1215.67(1+z_{abs}) \leq
\lambda_{max}$ was accepted only if its Lyman $\alpha$ line was present
(the spectrum ranges from $\lambda_{min}$ to $\lambda_{max}$ and $z_{abs}$ is
the redshift of the possible absorption system).

\item In the case of reasonably unblended multiplets, their line ratios should
be consistent with the expected values.

\item Every system with a column density for any ion larger than the one
for the associated Lyman $\alpha$  line was rejected.

\end{enumerate}

These criteria effectively ensured the rejection of many of the candidate
heavy-element absorption systems.   With all of the selected systems, we
used a more complete list of atomic transitions (in fact a list that contains
all transitions ever detected in QSO absorption system work) to search for
extra coincidences in the spectrum at the absorption system redshift.

The identification of heavy-element systems is particularly difficult in the
spectrum of QSO1055+021, because of the small spectral coverage. In this and
the other objects we have accepted as a heavy-element system anyone that
fulfils the above criteria, even if some of the lines are blended with
absorption lines from other systems, when the reality of such metal system
casts some doubts. However, we have taken these systems as real, since we
prefer to get the \lya\ line lists as clean as possible from any
contamination. To make sure that this decission will not affect the main goal
of this work, we checked that the wavelengths at which these doubtful systems
appear are far from the spectral regions in which the proximity effect from
the foreground QSO will be important, which is fortunately the case.

This is the (perhaps overcomplete) final list of heavy-element systems in the
spectra. There are previously reported spectra and line lists for QSO1228+077
(Robertson \& Shaver 1983) and QSO1222+228 (Bahcall, Osmer \& Schmidt 1969;
Morton \& Morton 1972; Bechtold et al. 1984) that have lower spectral
resolution.  We also compare our findings with these works.

\subsection{QSO1055+021}

\begin{enumerate}[(5)]
 \renewcommand{\theenumi}{(\roman{enumi})}

\item $z=2.14469 \pm 0.00003$. The only important observable lines
within the wavelength range are SiII $\lambda\lambda 1260,1304$. Both are
detected, but the second one is blended.

\item $z=2.32935 \pm 0.00002$. Observed Lyman $\alpha$ and SiIII
$\lambda 1206$ lines. No other line is detected, although SiII
$\lambda\lambda 1260,1304$ fall within the observed range.

\item $z=1.14360 \pm 0.00011$. The only lines within the observed spectral
range are  AlIII  $\lambda\lambda 1854,1862$; and both are detected and in good
agreement with theoretical line ratio although the second one is blended.

\item $z=0.31469 \pm 0.00003$. This is the most dubious system for this
QSO. The only available lines are AlI $\lambda\lambda 3083,3093$. Both are
detected, and the line ratio is correct,
but the second one blends slightly with another Lyman $\alpha$ line.

\item $z=1.37569 \pm 0.00007$. The red spectrum suggests the possibility of
another metal system, at $z=1.37569 \pm 0.00007$. This is based upon a
possible CaII doublet $\lambda\lambda 3934,3969$. No other line with this
redshift could be detected in the available wavelength range (both blue and
red arms) to confirm or reject this possibility, as we do not trust the
absence of AlII $\lambda 1671$ in the blue arm spectrum as a safe test to
reject this system.

\end{enumerate}

\subsection{QSO1222+228}

\begin{enumerate}[(10)]
 \renewcommand{\theenumi}{(\roman{enumi})}

\item $z=1.93724 \pm 0.00007$. This is a really interesting system.  Its
existence has been first reported by Bahcall et al. (1969).  In addition
Morton \& Morton (1972) also reported  the presence of structure in the
absorption lines. We can confirm the presence of three subsystems,  with
redshifts 1.93606 $\pm$ 0.00013 (A), 1.93724 $\pm$ 0.00007 (B) and  1.93853
$\pm$ 0.00007 (C), with a high degree of confidence: Lyman $\alpha$  lines,
together with the doublet NV $\lambda\lambda 1238,1242$ are seen  for all
three subsystems, and the line SiIII $\lambda 1206$ has been  fitted for
subsystems B and C. We do not believe that  the lack of SiII $\lambda\lambda
1190,1193$ is a relevant  problem. The computed $\Delta v$  for subsystems A
and C with respect to subsystem B are $121 \pm 39$ and $132 \pm 29  \, \kms$.

\item $z=1.23238 \pm 0.00005$. The CIV $\lambda\lambda 1548,1550$ doublet  is
present (the second component is blended). We have also detected a line of  CI
$\lambda 1560$ and FeII $\lambda 1608$. The only remaining
line in the  range, SiII $\lambda 1527$, is not detected.

\item $z=2.03669 \pm 0.00005$. Lyman $\alpha$ and SiII $\lambda\lambda
1190,1193$ are seen, although the second line of this doublet is blended. SiIII
$\lambda 1206$ is absent.

\item $z=1.51408 \pm 0.00006$. SiIV $\lambda\lambda 1394,1403$ and CII
$\lambda 1335$ are detected. No SiII $\lambda 1304$ absorption line is seen.

\item $z=0.47372 \pm 0.00003$. Three lines of the FeII multiplet
$\lambda\lambda 2344, 2375, 2383$ are seen.  No other line is --nor should
be-- observed in  this wavelength range.

\item $z=0.40722 \pm 0.00011$. Another system characterised by four FeII
lines. In this case, wavelengths are $\lambda\lambda 2344, 2383, 2587 and
2600$.

\item $z=0.23807 \pm 0.00004$. This is a MgII $\lambda\lambda 2796,2803$
doublet.  The first component is blended. MgI $\lambda 2853$ is not seen.

\item $z=0.26986 \pm 0.00003$. Another MgII doublet. The first component is
slightly blended. Two FeII lines and MgI $\lambda 2853$ are absent.

\item $z=0.29962 \pm 0.00005$. Yet another proposed MgII doublet. The
second component is blended. FeII lines are not detected.

\item $z=1.92307 \pm 0.00004$. Dubious SiII $\lambda\lambda
1190,1193$ and clear SiII $\lambda 1260$ line with correct line
ratios.  The associated Lyman $\alpha$ line is slightly blueshifted
($30\, \kms$ away) but is still consistent given the spectral resolution.

\end{enumerate}

	In comparison with previously available lists, we can say that the
z=1.93727 system has been totally confirmed, as well as its structure. We can
also report the detection of a Lyman $\alpha$ line at z=1.98095 $\pm$ 0.00003,
that matches the CIV doublet reported by Sargent et al. (1988) at z=1.9805
--also noticed as Lyman Limit system by Bechtold et al. (1984).

\subsection{QSO1228+077}

\begin{enumerate}[(5)]
 \renewcommand{\theenumi}{(\roman{enumi})}

\item $z=1.89678 \pm 0.00005$. This is a very clear heavy-element  system.  We
have detected its Lyman $\alpha$ line, SiII
$\lambda\lambda 1190,1193$,  SiII $\lambda\lambda 1260,1304$, and
SiIII $\lambda 1206$. This means that all important lines in our wavelength
range have been detected.

\item $z=2.01967 \pm 0.00006$. Detected Lyman $\alpha$ absorption,
together with SiIII $\lambda 1206$. There is no sign of the SiII
$\lambda\lambda 1190, 1193$ and 1260 lines.

\item $z=2.13687 \pm 0.00004$. As in (ii), we detect
Lyman $\alpha$ and SiIII $\lambda 1206$. SiII $\lambda\lambda
1190$ and 1193 are absent.

\item $z=2.29915 \pm 0.00013$. A dubious system, marked by the presence of
\lyb\ and CII $\lambda 1036$ absorption. There is no other line in
this region of the spectrum to aid a decission.

\item $z=1.20687 \pm 0.00007$. The CIV $\lambda\lambda 1548,1551$ doublet
is detected. The line ratio is correct, and the lines are only slightly
blended.

\end{enumerate}

The first three listed absorption systems of this QSO correspond to
systems A, B and C by Robertson \& Shaver (1983). Thus, we can confirm the
presence of B and C, which Robertson \& Shaver (1983) left as `probable'.

A very strong line in the spectrum of Robertson \& Shaver (1983)
at 4010.4 \AA is likely to correspond to the \lya\ line associated with the
system (iv) at $z=2.29915$.

\section{The Lyman $\alpha$ lines}

All the absorption lines not belonging to any of the above heavy-element
systems and falling in the QSO \lya\ forest are identified as \lya\
absorption lines. From the process presented in the previous section we can
conclude that the contamination from lines belonging to heavy-element
absorption systems is, if it exists at all, minimal. Since for all of the
absorption lines we have the column density, the Doppler dispersion parameter
and the redshift (together with the corresponding error estimates) we can in
principle obtain the distribution of these quantities. Some of them will, in
fact, be needed to properly model the strength of the proximity effect
produced by the foreground QSO.

\subsection{The column density distribution}

A first inspection of the column density distribution of the
\lya\ absorption systems for the three QSOs reveals that our lists are complete
down to (at least) $N_{HI}\approx 7\times 10^{13}\, \pcm$ in the case of the
QSOs 1222+228 and 1228+077, while we believe $N_{HI}\approx 10^{13}\, \pcm$ to
be our completeness limit for the improved resolution better signal-to-noise
spectrum of QSO1055+021. Although all of the analysis that we present here
is based on a column-density-limited sample, we can use our $N_{HI}$ cut-off
together with a typical value for $b$ ($35\ \kms $) and derive an equivalent
width cut-off of $\simeq $ 0.25 \AA. We remark that this limit is merely
orientative and does not imply any use of the equivalent widths of the lines
in the rest of this work.

% Stuff here table with results from fits to lya forest

\begin{table*}
  \begin{minipage}{150mm}
  \caption{Fits to the \lya\ absorption lines distribution.}
  \label{tab-fits}
  \begin{tabular}{ l  c c c c c c }\hline
OBJECT & $N^{inf}_{HI}$\footnote{Lower column density limit used in
calculations.}& $N^{sup}_{HI}$\footnote{Upper column density limit used in
calculations.}& $\beta$ & $\overline{b}$ &
 $\sigma_b$ & $A$ \\ \hline
1055+021       & $1.10^{13}$ & $\infty$ & $1.77^{+0.32}_{-0.25}$ &
 $26.5^{+3.5}_{-3.5}$   & $15.5^{+2.5}_{-2.5}$ & $10.7 \pm 2.5$ \\
1222+228       & $7.10^{13}$ & $\infty$ & $1.77^{+0.17}_{-0.15}$ &
 $48.5^{+12.5}_{-12.5}$ & $30^{+11}_{-7}$      & $ 7.0 \pm 1.6$ \\
1228+077       & $7.10^{13}$ & $\infty$ & $1.70^{+0.13}_{-0.11}$ &
 $70^{+18}_{-18}$       & $55^{+15}_{-9}$      & $ 8.6 \pm 1.6$ \\
Sample Average & $7.10^{13}$ & $\infty$ & $1.70^{+0.08}_{-0.09}$ &
           ---          &           ---        & $ 8.5 \pm 1.0$ \\ \hline
  \end{tabular}
  \end{minipage}
\end{table*}

As usual, a single power law has been fitted to the column density
distribution
\begin{equation}
p_N(N_{HI})\propto N_{HI}^{-\beta},
\end{equation}
via maximisation of the likelihood function which does not require the
binning up of the data. In the case of QSO1055+021, for which we have a
relatively small spectral coverage, there are two lines with large column
density ($N_{HI}>8\times 10^{15}\, \pcm$) for which we have no way of
testing whether or not they have associated metals. Since we have strong
suspicions that they could in fact belong to heavy-element systems, we prefer
to fit the column density distribution for this QSO up to a maximum column
density $N_{HI}<8\times 10^{15}\, \pcm$.

Fitting the column density distribution down to $N_{HI}>7\times 10^{13}\,
\pcm$ to the absorption lines of each QSO separately produces results that
are consistent with  the  standard value usually found in the literature
($\beta\approx 1.7$; see Table 6 for details). It is also worth mentioning
that the column density distribution for the lines in QSO1055+021 reveals a
flatter slope when the fit to the data from this QSO is performed down to
$N_{HI}=10^{13}\, \pcm$ ($\beta=1.55_{+0.13}^{-0.12}$). This might be a
result of the fact that the $p_N(N_{HI})$ flattens below some column
density, as it was reported by Carswell et al. (1987) and also indirectly
inferred by Webb et al. (1992).

By fitting the whole sample together the maximum likelihood method yields
$\beta=1.70_{-0.09}^{+0.08}$. This is the value that will be used in Section 5.

\subsection{The distribution of Doppler dispersion parameters}

As usual we have fitted a Gaussian form to the distribution of the $b$
parameter
\begin{equation}
p_b(b)=\gauss
\end{equation}
Because of the  lower resolution and signal to noise ratio, data from
QSO1222+228 and 1228+077 have a poorer definition in the measurements of
the Doppler  dispersion parameter $b$. To avoid contamination from badly
determined $b$s, we have removed from the  fitting procedure all lines with
$b$ determined with an error larger than $100\, \kms$.

The results are listed in Table 6. The lower spectral resolution data
(QSO1222+228 and 1228+077) produce larger $\overline{b}$ values and
dispersions than expected.  This difference can be understood as an artifact of
the resolution itself -- it is very difficult to detect lines much narrower
than the resolution.  When it is low, as in the QSO1222+228 and 1228+077
cases, the fitted values reflect more the spectral resolution itself than the
intrinsic distribution of the line Doppler dispersions. This is also reflected
in the error estimates of the $\bar b$ and $\sigma_b$ parameters, which have
been obtained by using the likelihood function as a probability density in
parameter space.  These errors are clearly larger for the low resolution data,
which implies a poorer determination, caused by line blending in the spectra.

\subsection{The redshift distribution}

The redshift distribution of lines, $\dndz$, is also required to model the
proximity effect. Our idea is basically to use the standard power-law
distribution (equation 1), and correct it according to the relative influence
of the background and foreground QSOs. But to do this, we must first make
sure that our sets of lines are compatible with the coefficients that we
will use, $A$ and $\gamma$. For example, Webb et al. (1992), use $A_o=11.6$
and $\gamma_{o}=1.987$ for lines with column density greater than $7\times
10^{13}\, \pcm$ which appears to be appropriate for intermediate to high
resolution studies.

Unfortunately, it is not possible to perform an accurate $\gamma$ fitting in
such a short redshift range and with such restricted wavelength coverage in
each spectrum. For this reason we computed the value of $A$ for each QSO
taking  $\gamma =\gamma_{o}$ fixed. The results are also listed in Table 6,
together with an error estimated in terms of Poisson counting statistics
over the redshift range covered. The values of $A$ are consistent with the
$A_o$ value quoted in the literature. The overall value for the normalisation
is $A=8.5\pm 1$, which is the value that we shall use in the next Section.

\section{Photoionization of the Lyman $\alpha$ forest by a foreground QSO}

\subsection{Modelling the Proximity Effect by a nearby QSO}

As explained in the Introduction, the inverse effect (in our case the proximity
effect) is most likely as result of an enhancement of
the UV ionizing field in the neighbourhood of a QSO which consequently
increases
the ionization fraction of the \lya\ clouds.  Although other models with
similar philosophy may also explain the inverse effect (for example
gravitational infall of the \lya\ clouds onto the QSO), we have modelled the
proximity effect in terms of photoionization of low-density highly ionized
clouds of gas -- much similar to the way that BDO modelled the inverse effect
but including the ionization from the foreground QSO.

The basic idea is that for a highly ionized plasma, where collisional
ionization can be ignored, the neutral gas density is inversely proportional to
the total UV ionizing intensity. If $J_0$ is the general UV background
intensity (in $\erg$) at the local Lyman limit frequency, and $F_Q$ is the UV
flux (in ${\rm erg}\, {\rm s}^{-1}\, {\rm cm}^{-2}\, {\rm Hz}^{-1}$) produced
by a QSO located at a distance $d_Q$ from the \lya\ cloud, the HI column
density will be reduced by the presence of the QSO according to
\begin{equation}
N_{HI}\propto(4\pi J_0+F_Q)^{-1}\propto (1+\omega_Q)^{-1}.
\end{equation}
where
\begin{equation}
\omega_Q=\frac{F_Q}{4\pi J_0}.
\end{equation}

The flux of the QSO on the \lya\ cloud is just $L_Q/(4\pi d_Q^2)$ where
$L_Q$ is the QSO luminosity at the local Lyman limit frequency and $d_Q$ is
the distance from the QSO to the \lya\ cloud. We have estimated $L_Q$ in
the same way as Tytler (1987), by using the measured V magnitude and a
suitable $K$ correction. We have used the $K$ correction presented by Evans
\& Hart (1977), assuming that it holds up to a redshift of 2.7. We have
also assumed that the emission from the foreground QSO is isotropic (i.e.,
no beaming) and that the measured magnitude has not been seriously
contaminated by gravitational lensing or micro-lensing amplification along
the line of sight. Another implicit assumption in this framework is that
the foreground QSO luminosity is not variable. We will discuss this point
further in Section 6.

Within this simple framework, the net effect of the presence of a nearby
QSO would be to {\it reduce} the column densities of \lya\ systems. Since
we are working with column-density-limited samples, some of the clouds
that would be detectable in the absence of the proximity effect will now
get a column density below our thresholds and they will therefore disappear.
Now the number of absorption lines (for a fixed $\omega_Q$) above the
completeness threshold will decrease by a factor $(1+\omega_Q)^{1-\beta}$.
Therefore the redshift distribution given by equation (1) has to be revised
by the presence of both the background and the foreground QSO, i.e.,
\begin{equation}
\dndz = A\, (1+z)^{\gamma}\, (1+\omega_{f})^{1-\beta}
\, (1+\omega_{b})^{1-\beta}
\end{equation}
for each QSO distribution.

Notice that if we adopt the values $A=8.5$, $\gamma=1.987$ and the
estimates of the QSO luminosities as discussed above, the only free
parameter in this equation is the UV background $J_0$.  In what
follows we shall assume that $J_0$ does not depend on redshift, given
the relatively restricted redshift range spanned by our observations.
In fact, models for the origin of the UV background in terms of the
integrated QSO emission produce a UV background which does not vary
very much in the range under consideration.  This is a result of the
cancellation of the expansion of the universe by the peak in the
redshift distribution of the QSO population (see, e.g., Atwood,
Baldwin \& Carswell 1985; Bechtold et al. 1987).

Since we are interested in the proximity effect caused by the foreground
QSOs, we will ignore the inverse effect portion of the different spectra,
i.e., the region where photoionization by the background QSO will affect
the distribution of lines.  This is only relevant to the spectrum of
QSO1222+228 in which we have ignored a range of $5000\, \kms$ around the
redshift of the QSO. We checked with our best-fitting values for $J_0$
that this region was large enough to exclude the inverse effect region of
this QSO. This also allows us to ignore the $(1+\omega_b)^{1-\beta}$
factor in eq (7).

\subsection{Estimates of the UV ionizing flux}

We have performed a one-parameter maximum-likelihood fit to $\dndz$ to find
$J_0$. The absence of proximity effect would yield $J_0=\infty$.   The value of
$J_0$ that maximizes the likelihood function (see Fig. 4) is  $\log J_0=-20.5$
in the usual units ($\erg$).

%\begin{figure}
%\vspace{7cm}
%\label{fig-lik}
%\caption{Likelihood function in terms of $\log J_0$. The inverse effect region
%in QSO1222+228 has been excluded.}
%\end{figure}

%\begin{figure}
%\vspace {6cm}
%\caption{Normalised \lya\ absorption line density (after removal of the
%%overall
%trend with $z$) as a function of the flux from the foreground QSO}
%\label{fig-bins}
%\end{figure}

In principle, this kind of fit should also allow us to estimate the  errors
in the fitted parameter, by using the likelihood function as a probability
in parameter space. However, it appears that the likelihood function is
unable to be normalized, because once the UV flux $J_0$ has gone above some
value the fitting is insensitive to it (the foreground QSO no longer
influences the \lya\ forest) and the Likelihood function goes to a non-zero
constant as $J_0\to \infty$.  Another way to estimate errors from the
likelihood function (used, e.g., by Kulkarni \& Fall 1993 to estimate the
UV  background at low $z$ via the inverse effect) uses the derivatives of
the  likelihood function around its maximum.  However, this is not feasible
in our  case as the peak that we obtain in the likelihood function is not
very pronounced. On the other hand (see Fig. 4), the Likelihood function
presents a  strong decay for low values of $\log (J_{0})$, which would mean
the rejection of very low values of $J_0$ as the proximity effect should be
much stronger than observed.

In order to check whether our result is significant (i.e., whether a
proximity effect is detected) we have carried out a number of tests.
Unfortunately, none of them places our measurement more than $1\sigma$ away
from the inexistence of the proximity effect ($J_0=\infty$).  The most
illustrative method is to generate synthetic line lists for the three QSOs
in our sample with line densities according to equation (6) and then apply
the same method to the simulated lists as to the real data.  The result,
rather discouraging, is that with the current sample and with $J_0=\infty$
there is a probability $\sim 35$ per cent of measuring a value of
$J_0<10^{-20.5}\erg$ by chance.

A complementary test has been provided by the use of the bootstrap method,
as presented by Efron \& Tibshirani (1986). By using 500 bootstrap
reshuffled absorption line samples, we have seen that only 15 per cent of
them do not show any local maximum in the likelihood function (i.e., only 15
per cent of the bootstrapped samples give $J_0=\infty$). We can now use this
distribution of $J_0$ values to try to assess error bars to the value that
maximizes the likelihood function. The result is that the $1\sigma$ upper
bound just hits the $J_0=\infty$ value in complete agreement with the previous
method.

We then conclude that although our results are consistent with a proximity
effect caused by a UV background of the order of $J_0=10^{-20.5}\, \erg$,
we cannot rule out that the effect is in fact not present. In Figure 5 we
show a binned distribution of the \lya\ absorption lines as a function of
$\log (F_{f}^{2}/4\pi)$ --where $F_f$ is the flux from the foreground QSO,
locally normalised to the overall redshift distribution (equation 1). The
proximity effect implies a decrease of this function at large values of the
foreground QSO flux. We also show the expected distribution  for different
values of $J_0$.  This diagram illustrates that although there is some hint
that the proximity effect might be present, its detection (i.e.,
discrimination from the $J_0=\infty$ curve) is indeed difficult.

Unfortunately, the closest distance from a foreground QSO to the line of
sight of the corresponding background QSO is quite large ($1.7\,
h_{50}^{-1}\,  {\rm Mpc}$) which does not enable us to explore the diagram
presented in Fig. 5 at higher values of $\log (F_{f}^{2}/4\pi)$. Indeed,
this is much better achieved by including the inverse effect caused by the
background QSO itself. In fact, and as a test,  we have used the whole
available spectrum of QSO1222+228, which also contains the inverse effect
range. The likelihood function now exhibits a very pronounced peak (see
Fig. 6) and the measured UV background is now $\log J_0=-21.1 \pm 0.6 \erg$.

We can also use the bootstrap results to find lower bounds to the UV
intensity within the photoionization model. These come from the fact that a
UV background that is too low would imply an unacceptably large proximity
effect. At  95 per cent confidence, the UV ionizing flux is $J_0>10^{-21.8}\,
\erg$. This limit is quite secure because our sample contains a pair
(QSO1055+021) that would show some hint of proximity effect if $J_0$ were low
enough, while no evidence is found for it in that particular spectrum. We
discuss in next Section the implications of this lower limit for the origin
of the UV background at high redshift.

%\begin{figure}
%\vspace{7cm}
%\label{fig-lik2}
%\caption{Likelihood function in terms of $\log J_0$. The inverse effect region
%in QSO1222+228 has not been excluded.}
%\end{figure}

\section{DISCUSSION}

To our knowledge there are only two previous observational works about
the detection of the proximity effect by foreground QSOs (Crotts 1989;
Dobrzycki \& Bechtold 1991). Crotts(1989) uses spectra of QSO1623+268 and
three other QSOs in its neighbourhood. In that work no proximity effect is
detected at 2.4$\sigma$ level, and the author suggests that either the QSO
radiation is highly anisotropic or that the proximity effect has causes
other than radiation. Dobrzycki \& Bechtold (1991) find a void at $z=3.17$
towards QSO0302-003, which is not coincidental with the redshift of a known
foreground QSO ($z=3.223$). In order to interpret this void in terms of the
proximity effect of the foreground QSO, these authors conclude that its
radiation has to be highly anisotropic.

Like all of the above authors, we have ignored the fact that there is
possibly variability in the QSOs. There are two possible sources for this
variability. First, we could be observing some kind of long-term fading in
the QSOs. We do not believe this effect to be significant, since pure
luminosity evolution models for the QSO population show that the time-scale
for this effect is much longer than the light travel time from the QSO to
the cloud (tipically 30 Myrs). On the other hand, we cannot forget the
possibility of having short-scale variability. In any case, it is impossible
to model this effect, as there is only statistical information on this
phenomenon. Moreover we can expect this effect to be less important when a
larger sample is used (because of cancellation of individual changes). Hence,
we believe that to neglect variability is our best choice. It is worth
mentioning that this problem is not present in the measurements of $J_{\nu}$
when the proximity effect of the QSO on its own line of sight is used.

We think that the set of statistical tools used in our work to search for
the proximity effect is quite appropriate. The use of simulations of both
the line lists and the bootstrap technique to assess significance levels and
confidence intervals are in very good agreement. Taking the Likelihood
function for the redshift distribution of the lines as the statistic to be
maximized in the fitting procedure, we do not require to bin the data, which
is very important when only a limited amount of data are available.  A {\it
log likelihood} analysis [as used, e.g., by Shafer (1983) in a different
context] also gave similar results.  This method also showed that the sample
should be much increased (by a factor of at least 10 in the number of lines)
in order to be able to exclude $J_0=\infty$ at, say, 95 per cent confidence.

We have also made other attempts to detect the proximity effect, with similar
success. For example, comparing the summed column density in a window of $\pm
5000\, \kms$ around the known redshift of the foreground QSO with the same
value for a set of simulations generated with different trial values of $J_0$,
also shows a best fit for $J_0\sim 10^{-20.5}\, \erg$; but again, when we try
to estimate the error bar for that value, we find that $J_0=\infty$ (i.e., no
proximity effect) is only about $1\sigma$ away.

Comparing our results with the available UV background flux models (basically
Bechtold et al. 1987 which are also used by  BDO), we can reach some
conclusions. We have selected  the models presented in that paper with
$q_0=0.5$, including absorption from Lyman $\alpha$ clouds, diffuse neutral
hydrogen and thick Lyman $\alpha$ absorbers.  In that work, the UV background
contributed by QSOs is estimated at different redshifts by assuming different
QSO evolution models. Since our study is aimed at testing this quantity in
the range $z\sim 1.7-2.7$, where most QSOs have been switched on, we cannot
distinguish between models with different types of redshift cut-offs. The
final result of these models is that the UV ionizing flux at the redshifts
under consideration lies between $10^{-21.8}\, \erg$ and $10^{-22}\, \erg$,
i.e. below our 2 $\sigma$ lower limit. Similar results are obtained when
comparing with models in Madau (1992) and Meiksin \& Madau (1993). This
means that our work agrees with the suggestion by Bechtold et al. (1987),
BDO and others, i.e. that {\it observed} QSOs do not appear to be either
luminous or numerous enough to explain all of the UV background flux that is
inferred at intermediate redshift.

One possible alternative is that some kind of absorber (dust?) is obscuring an
important part of the light from QSOs between redshift 2 and 0. This would
imply that the QSO population at medium to high redshift is larger than the one
we observe (see, e.g., models by Heisler \& Ostriker 1988). That, however,
would imply a systematic reddening of the high redshift QSOs which is not
consistent with the low dust content of the thick absorbers (Fall \& Pei 1989;
Fall, Pei \& MacMahon 1990).

Another possibility that could explain the UV excess would be to postulate
the presence of a large amount of unobserved high-redshift star-forming
galaxies (see e.g. Madau 1991). It is hard to make this thesis compatible
with present-day galaxy luminosity functions. In fact, there is much
controversy nowadays about the feasibility of the presence of a large number
of {\it dwarf galaxies} with $22 < M < 27$, at $z \simeq 2$ (see, e.g.,
Cowie, Songaila \& Hu 1991; Efstathiou 1992; Cole, Treyer \& Silk 1992).
None the less, one would expect that high-z star-forming galaxies contain
large amounts of dust (as is observed at low redshift), which would preclude
any relevant amount of UV photons escaping from these objects.

An explanation for the proximity effect which is also  entirely consistent with
our data is the one where QSOs reside in the depths of strong gravitational
potential wells and \lya\ clouds fall onto those potential wells.  This cleans
up the neighbourhood of a $z=2.5$ QSO within a  radius of $0.3 h_{50}^{-1}\,
\sigma_{100}\, {\rm Mpc}$ in less than a Hubble time, where $\sigma_{100}$ is
the velocity dispersion in the potential well in units of $100\, \kms$.  If
QSOs exist in high-density environments where velocity dispersions are high, it
is possible to explain the proximity effect in these terms.  In fact, Vishniac
\& Bust (1987) and Barcons \& Webb (1990) discussed similar mechanisms to
explain the lack of clustering of \lya\ clouds around galaxies (represented by
heavy-element systems) in biased scenarios for the formation of these objects,
where such clustering would be expected.

We believe that the results obtained by Crotts (1989) and Dobrzycki  \&
Bechtold (1991), are not inconsistent with the present work. Using a very
simple model for the QSO emission (an opaque torus surrounding the central
source as suggested by the AGN unified scheme) and the opening angle $\sim
140^{\circ}$ given  in Dobrzycki and  Bechtold (1991), it turns out that the
case where the system behaves {\em as  if the radiation from the foreground
QSO were emitted isotropically} (i.e. when the opacity towards the observer
and towards the point of closest approach in the line of sight to the
background QSO are similar)  has a probability $\simeq 50$ per cent. It is
also expected that in some cases ($\simeq 25$ per cent according to that
na\"\i ve model) the torus obscures the direction towards the background QSO
line of sight but not the direction towards the observer, and  the inverse
effect would not be seen. In the remaining $\simeq 25$ per cent of the cases
the torus will obscure the foreground QSO from the observer but would ionize
the  background QSO line of sight.  All these numbers are derived from the
estimate by Dobrzycki and  Bechtold (1991) of the obscuring angle, which is
based in a single case and therefore could change when more information
becomes available from other QSO pair observations.

The model for \lya\ clouds, where these are low-density highly ionized
galaxy-sized gas aggregates, provides the most basic scenario that is
consistent with our observations. Although this is the model preferred by
the data we cannot, unfortunately, rule out other non-local explanations
for the inverse effect. More definite conclusions, especially the ones
concerning the isotropy of the emission, could be achieved by means of
observing a larger sample of QSO pairs.

\vspace{.7cm}

\begin{bf}
\noindent Acknowledgments
\end{bf}

AFS thanks Adam Dobrzycki for useful
comments on an original version of this paper. AFS and RC acknowledge
support by a research grant by
the Spanish MEC.  Partial financial support for this work was provided by the
DGICYT under project PB92-0741 and by the Comission of the European Union under
the `Human Capital and Mobility' contract CHRX-CT92-0033. The Isaac Newton
Telescope (INT) and the William Herschel Telescope (WHT) are operated on the
island of La Palma by the Royal Greenwich Observatory in the Spanish
Observatorio del Roque de Los Muchachos of the Instituto de Astrof\'\i sica de
Canarias. The spectrum of 1305+298 was taken during a service night at WHT.

\bsp

\newpage

\begin{bf}
\noindent FIGURE CAPTIONS
\end{bf}

Figure 1: The spectrum of QSO 1055+021. See text for details.

Figure 2: The spectrum of QSO 1222+228. See text for details.

Figure 3: The spectrum of QSO 1228+077. See text for details.

Figure 4: Likelihood function in terms of $\log J_0$. The inverse effect
region in QSO1222+228 has been excluded.

Figure 5: Normalised \lya\ absorption line density (after removal of the
overall trend with $z$) as a function of the flux from the foreground QSO.

Figure 6: Likelihood function in terms of $\log J_0$. The inverse effect
region in QSO1222+228 has not been excluded.

\vspace{2cm}

\begin{bf}
\noindent TABLE CAPTIONS
\end{bf}

Table 3: QSO 1055+021: absorption lines list. See text for details.

Table 4: QSO 1222+228: absorption lines list. See text for details.

Table 5: QSO 1228+077: absorption lines list. See text for details.

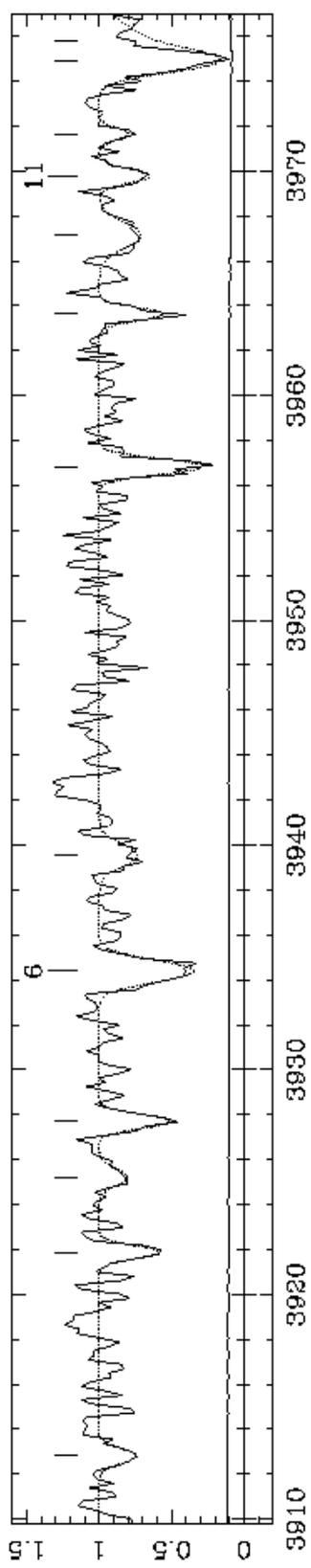
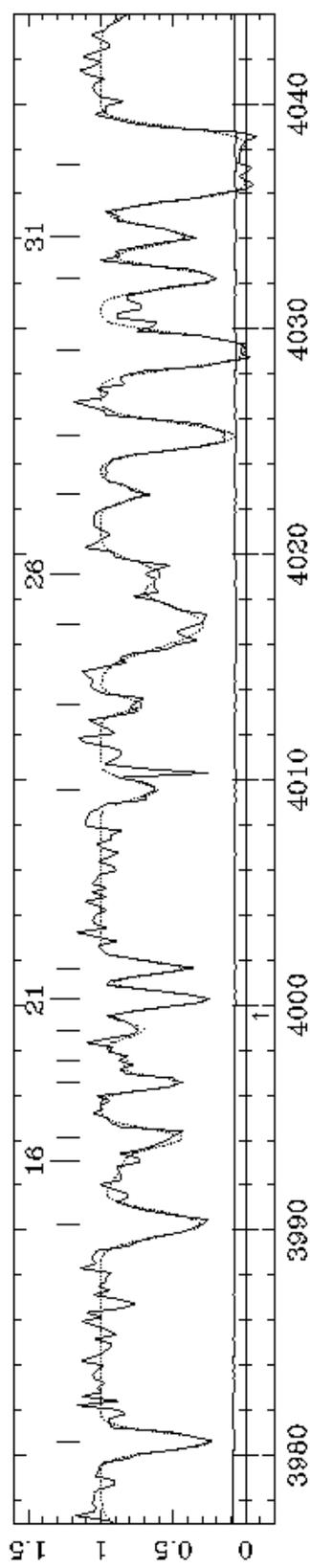
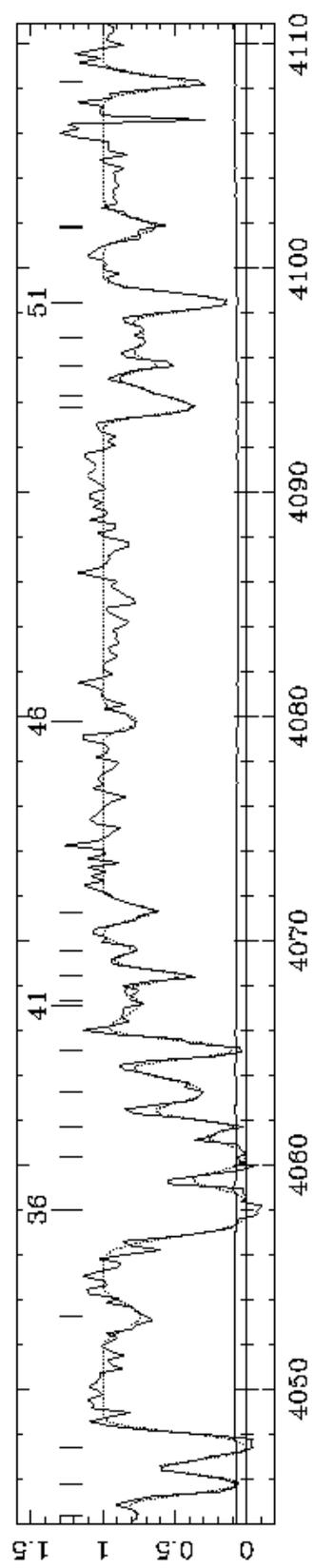
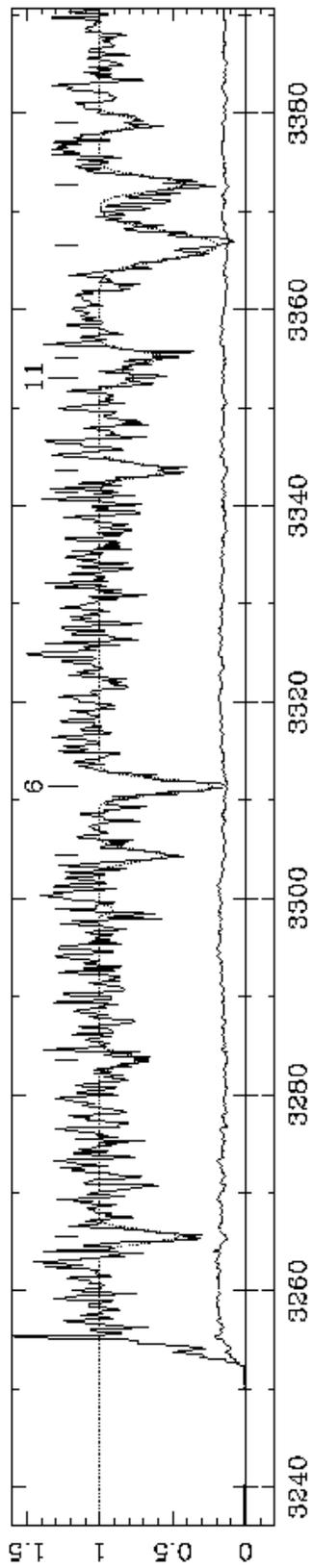
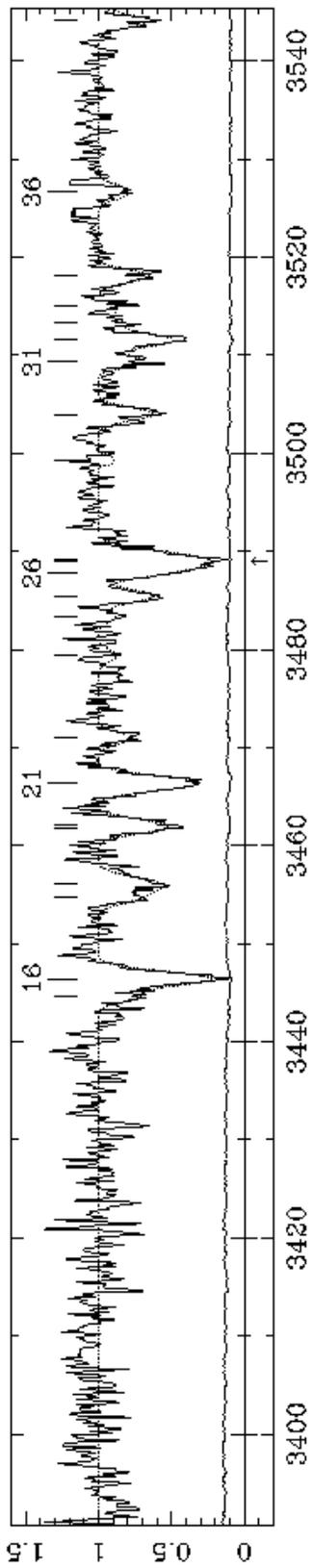
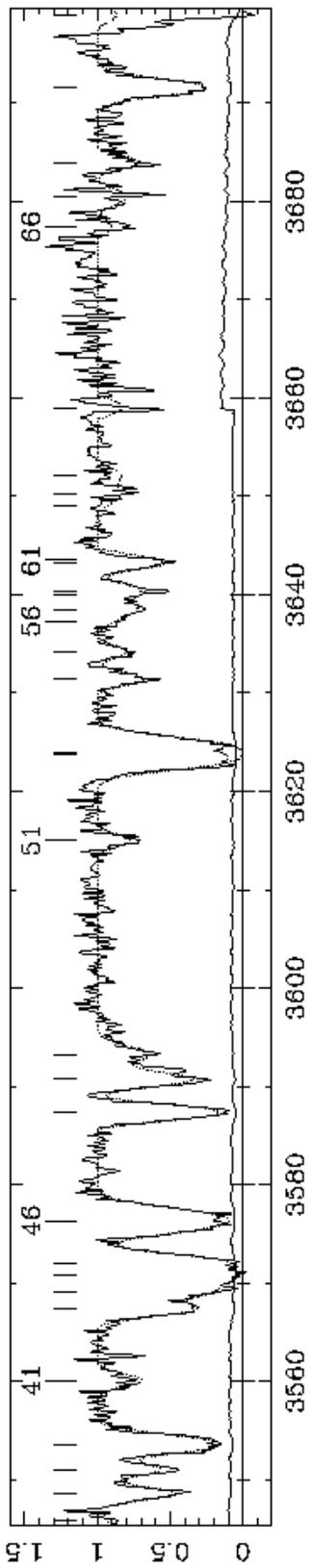
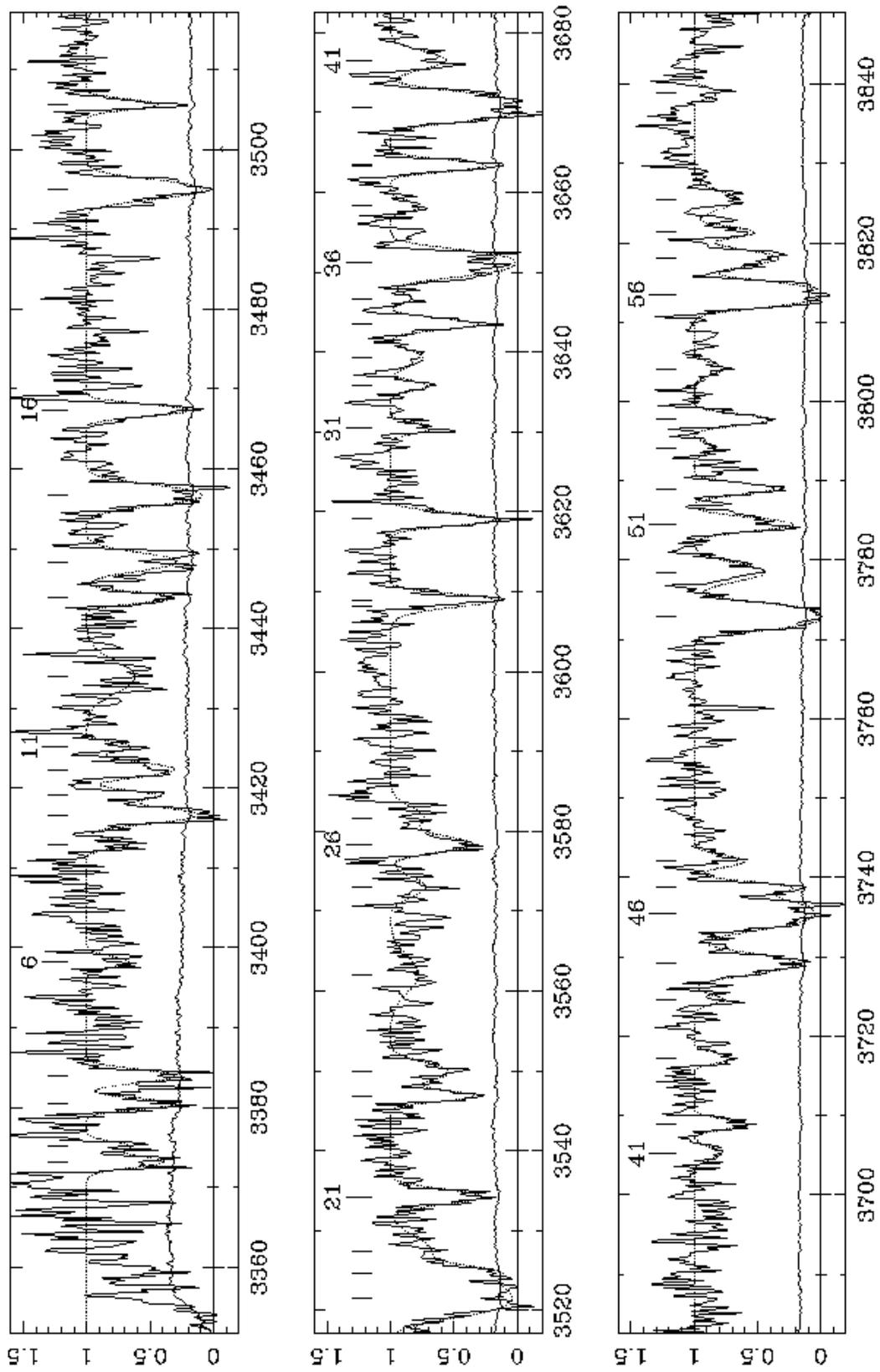
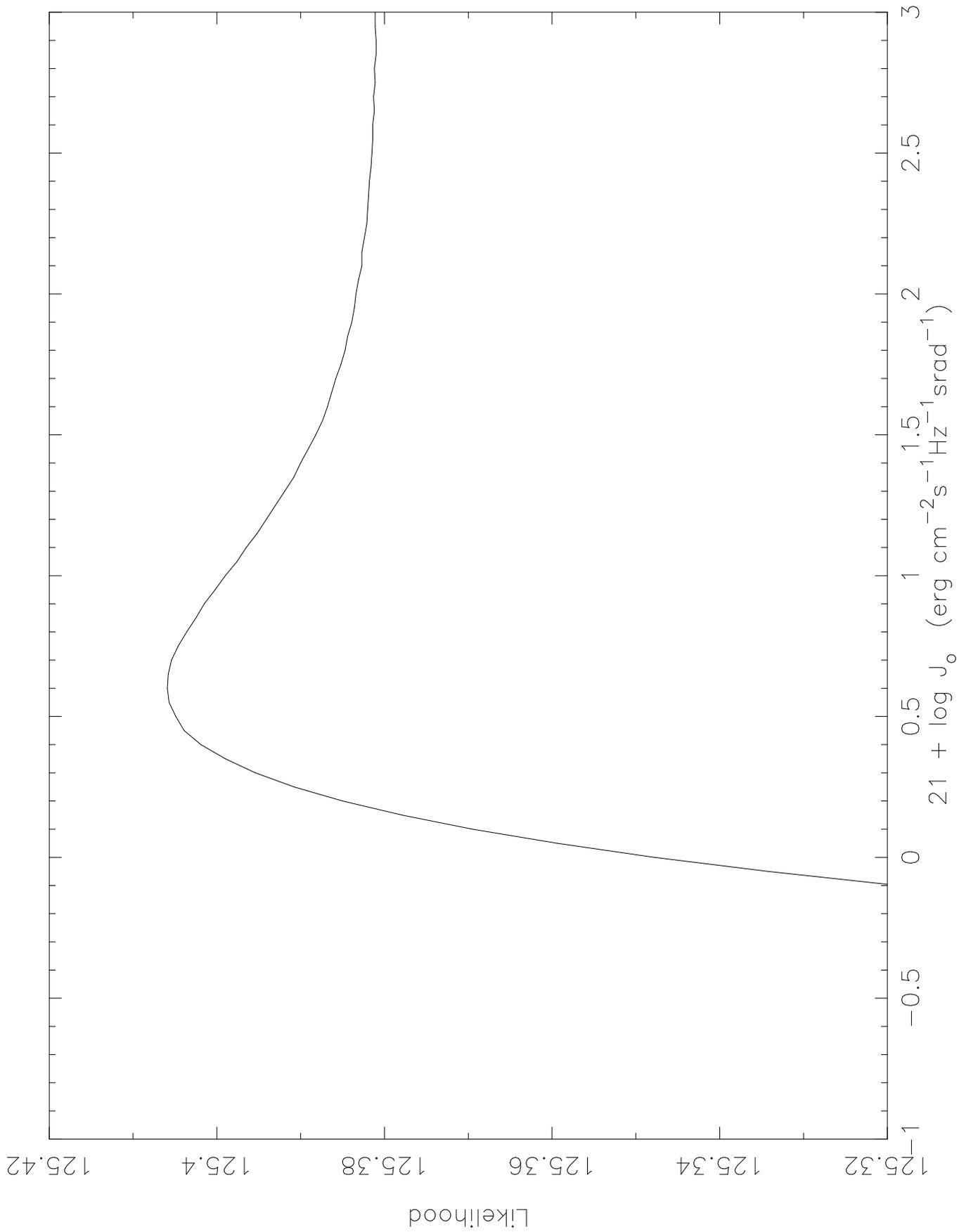
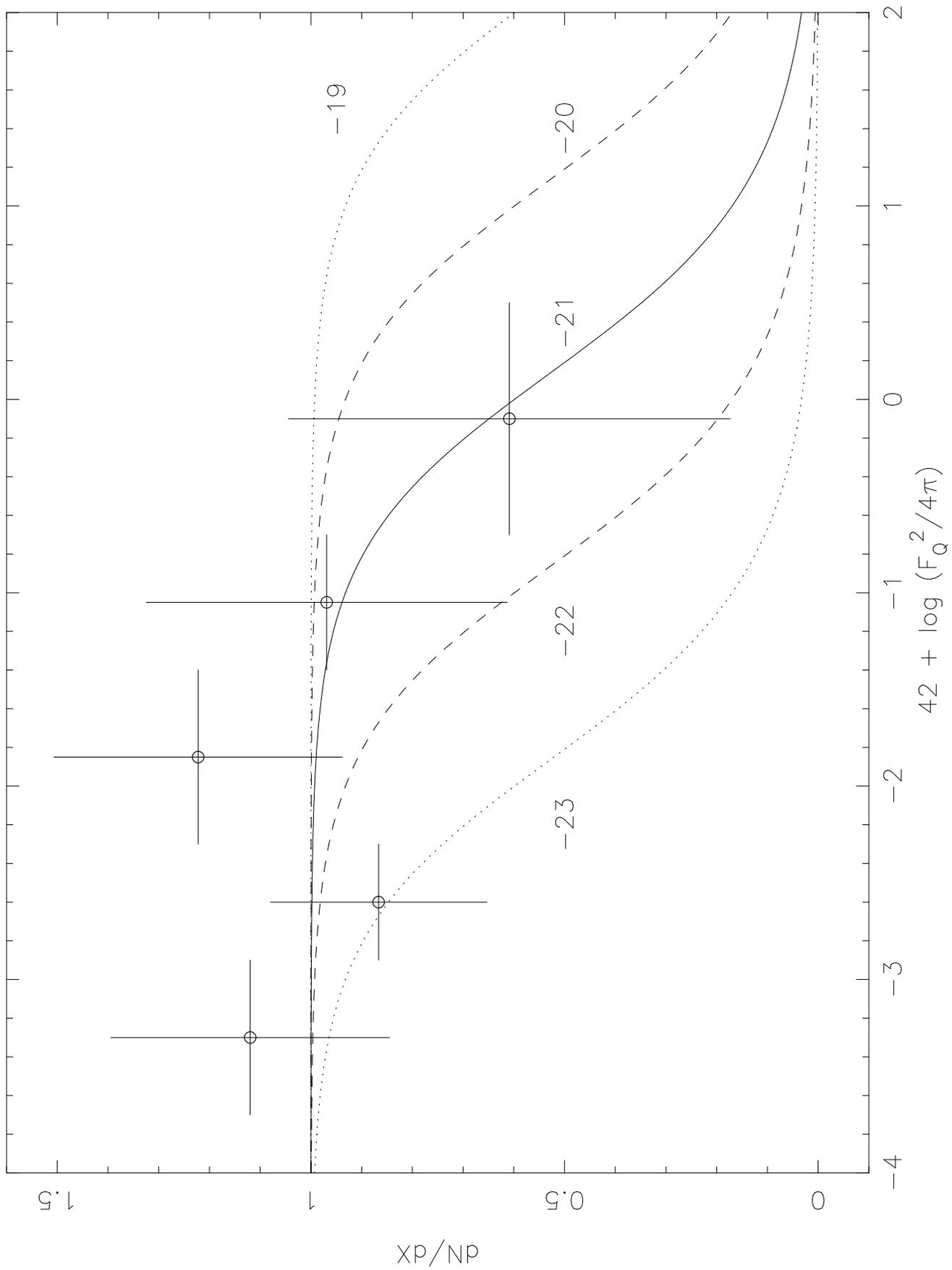
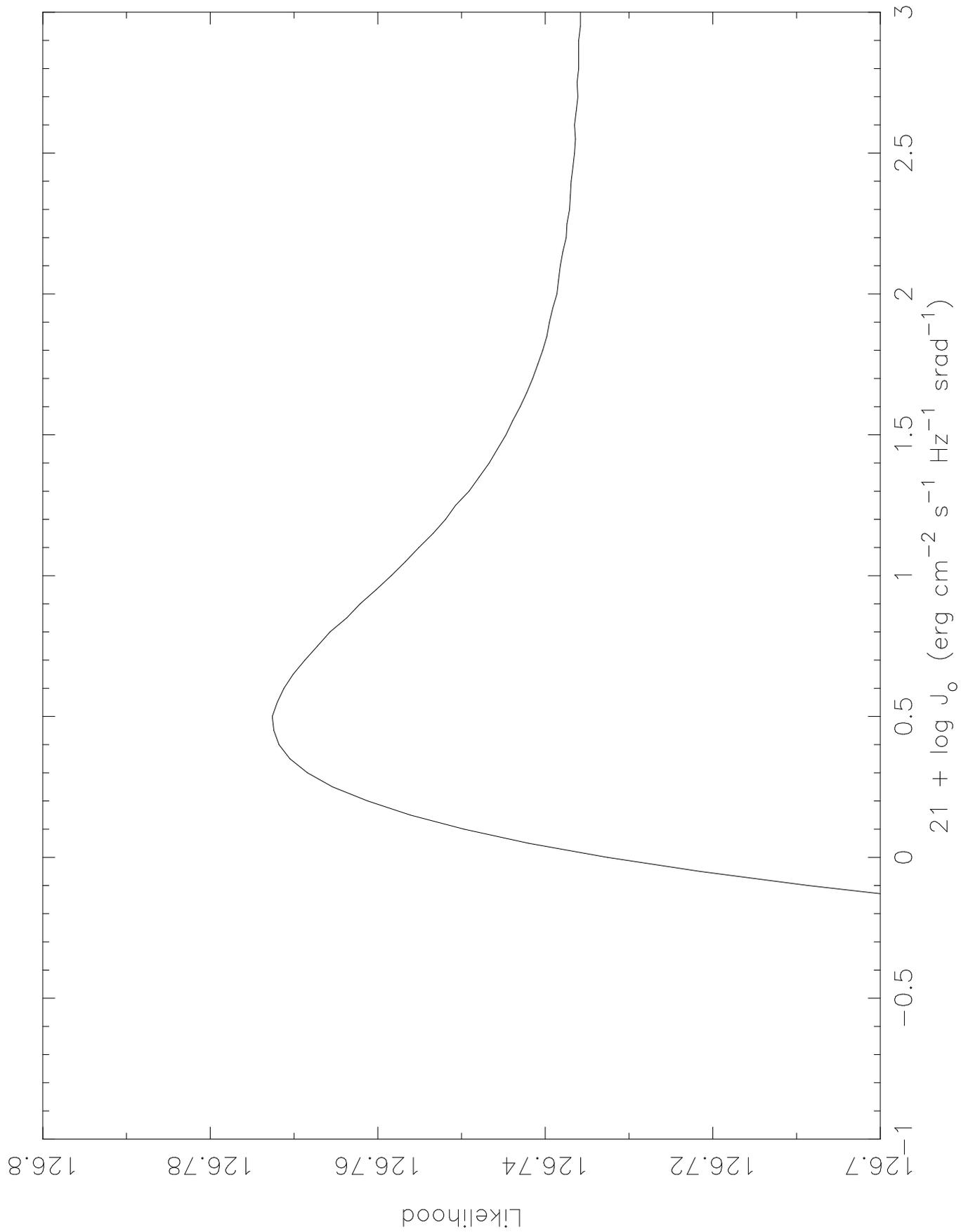
\end{document}